%% file: main_conf.tex
\documentclass[conference,10pt]{IEEEtran}
\usepackage{amsmath, mathrsfs, mathtools}
\usepackage{amsfonts}
\usepackage{amssymb}
\usepackage{color}
\usepackage{multirow}
\usepackage{stmaryrd}
\usepackage{yfonts}
\usepackage{mathabx}
\usepackage{caption}
\usepackage{subcaption}
\usepackage{float}
\usepackage{stfloats}
\usepackage{amsfonts}
\usepackage{tcolorbox}
\usepackage[numbers,sort&compress]{natbib}
\usepackage{multicol}
\usepackage{algorithm}
\usepackage{algpseudocode}
\usepackage{pifont}
\usepackage{varwidth}
\usepackage{afterpage}
\usepackage{balance}
\usepackage{lipsum}
    
\usepackage{tikz, pgfplots,graphicx,xcolor}
\usetikzlibrary{plotmarks,spy,backgrounds}
\pgfplotsset{compat=newest}

\input{symbols.tex}

\input{macros.tex}

\newcommand*{\Perm}[2]{{}^{#1}\!P_{#2}}

\allowdisplaybreaks
\newcommand{\normd}[1]{{\left\vert\kern-0.25ex\left\vert\kern-0.25ex\left\vert #1 
		\right\vert\kern-0.25ex\right\vert\kern-0.25ex\right\vert}}
\title{Cooperative Multistatic Target Detection in Cell-Free Communication Networks}
\author{Tianyu Yang, Shuangyang Li, Yi Song, Kangda Zhi, and Giuseppe Caire\\
	\vspace{2mm}
	$\{$tianyu.yang, shuangyang.li, yi.song, kangda.zhi, caire$\}$@tu-berlin.de
}

\begin{document}
\maketitle
\begin{abstract}
    In this work, we consider the target detection problem in a multistatic integrated sensing and communication (ISAC) scenario characterized by the cell-free MIMO communication network deployment, where multiple radio units (RUs) in the network cooperate with each other for the sensing task. By exploiting the angle resolution from multiple arrays deployed in the network and the delay resolution from the communication signals, i.e., orthogonal frequency division multiplexing (OFDM) signals, we formulate a cooperative sensing problem with coherent data fusion of multiple RUs' observations and propose a sparse Bayesian learning (SBL)-based method, where the global coordinates of target locations are directly detected. 
    Intensive numerical results indicate promising target detection performance of the proposed SBL-based method. Additionally, a theoretical analysis of the considered cooperative multistatic sensing task is provided using the pairwise error probability (PEP) analysis, which can be used to provide design insights, e.g., illumination and beam patterns, for the considered problem. 
    % {\color{red} here is a bit strange when you mention the omnidirectional beam is better than random beam. I think there should be a tradeff somewhere. Also, intuitively this suggest the marco diversity from cell-free is not enough, which sounds weird.}
    
    %numerically verifies the advantages of the equal-powered omnidirectional beam compared to the random beam direction.  
    
    %with orthogonal frequency division multiplexing (OFDM) signals, where the cooperative and coherent sensing task of multiple radio units (RU) is addressed. 
    
\end{abstract}


\section{Introduction}
Integrated sensing and communication (ISAC) has emerged as a potential solution to the spectrum crunch problem facing in modern wireless technologies~\cite{liu2022integrated}. ISAC allows the dual use of both communications and sensing at the same frequency bands, allowing similar hardware architectures for the coexistence of two functionalities. Due to its importance and promising potentials, ISAC has been listed as one of the primary usage scenarios in the forthcoming 6G network \cite{ITU2023}.

%As the carrier frequency of the communication system ever increases to a high level, the operation band for communication also becomes proper for the radar functionality. As a consequence, the concept of integrated sensing and communication (ISAC) emerged in recent years and has grown as a key technique in the coming 6G era, with the goal of fusing sensing with communication in the same physical layer \cite{liu2022integrated}. By sharing the same equipment, spectrum, and signals, ISAC enjoys lower costs, and higher spectral and energy efficiency compared to counterparts that require dedicated transceiver designs \cite{liu2022integrated, ahmadipour2022information}.

In this paper, we consider a communication-centric ISAC design in a network setup, where multiple radio units (RUs) cooperatively detect the target in a cell-free MIMO communication network. This problem is well-motivated by the recent research focus on ``network as a sensor'' \cite{liu2022networked}.
% {\color{red} cite paper "Networked Sensing in 6G Cellular Networks: Opportunities and Challenges"}. 
A realization of such networked sensing in ISAC regime is the cell-free MIMO systems, where a dense network of RUs is deployed.
In a cell-free network with multiple transceivers serving sensing tasks, cooperative sensing can potentially improve the sensing performance \cite{liang2011design,ahmadipour2022information}. An easy cooperation is realized by sharing independent monostatic sensing results from each sensor and a subsequent fusion is applied to improve the sensing accuracy. This work focuses on a much stronger cooperation that involves a direct central processing of all observations of the distributed multistatic sensor network.

%The simplest sensing scheme is monostatic sensing, where the transmitter and receiver are co-located. On the contrary, the bistatic sensing scheme deploys the transmitter and receiver in different positions avoiding challenging full-duplex operation. Furthermore, the networked multistatic sensing scheme involves multiple distributed transceivers that shape a sensing network for better sensing performance due to more diverse observations of the environment.  

Despite the clear benefit of strong cooperative network sensing, the coherent central process of the wideband signals from distributed arrays is difficult. In this work, we address such strong cooperative sensing problems in the wideband cell-free communication systems with typical communication signals, i.e., orthogonal frequency division multiplexing (OFDM) signals. Instead of separately estimating the delay (distance) and angle of the targets, e.g., in \cite{song2024compressed}, we directly estimate the locations of targets in the global coordinate system by coherently exploiting the angle-delay resolutions of distributed arrays and wideband signals. With an alternate illumination-receiving pattern of the distributed arrays, wideband response signals of multiple targets are centrally collected and coherently used for global estimation of target locations. Specifically, we apply a discretization of the space and formulate the cooperative sensing problem as a compressed sensing problem, where a grid-based sparse Bayesian learning (SBL) approach is proposed to estimate the radar cross section (RCS), which in turn gives the target locations in the grid. 
Moreover, a theoretic pairwise error probability (PEP)-based analysis is provided for the detection error of a matched model, which numerically verifies the comparison of two beam patterns under tested short-time observations.        

\begin{figure}[!t]
        \centering
        \includegraphics[width=.8\columnwidth]{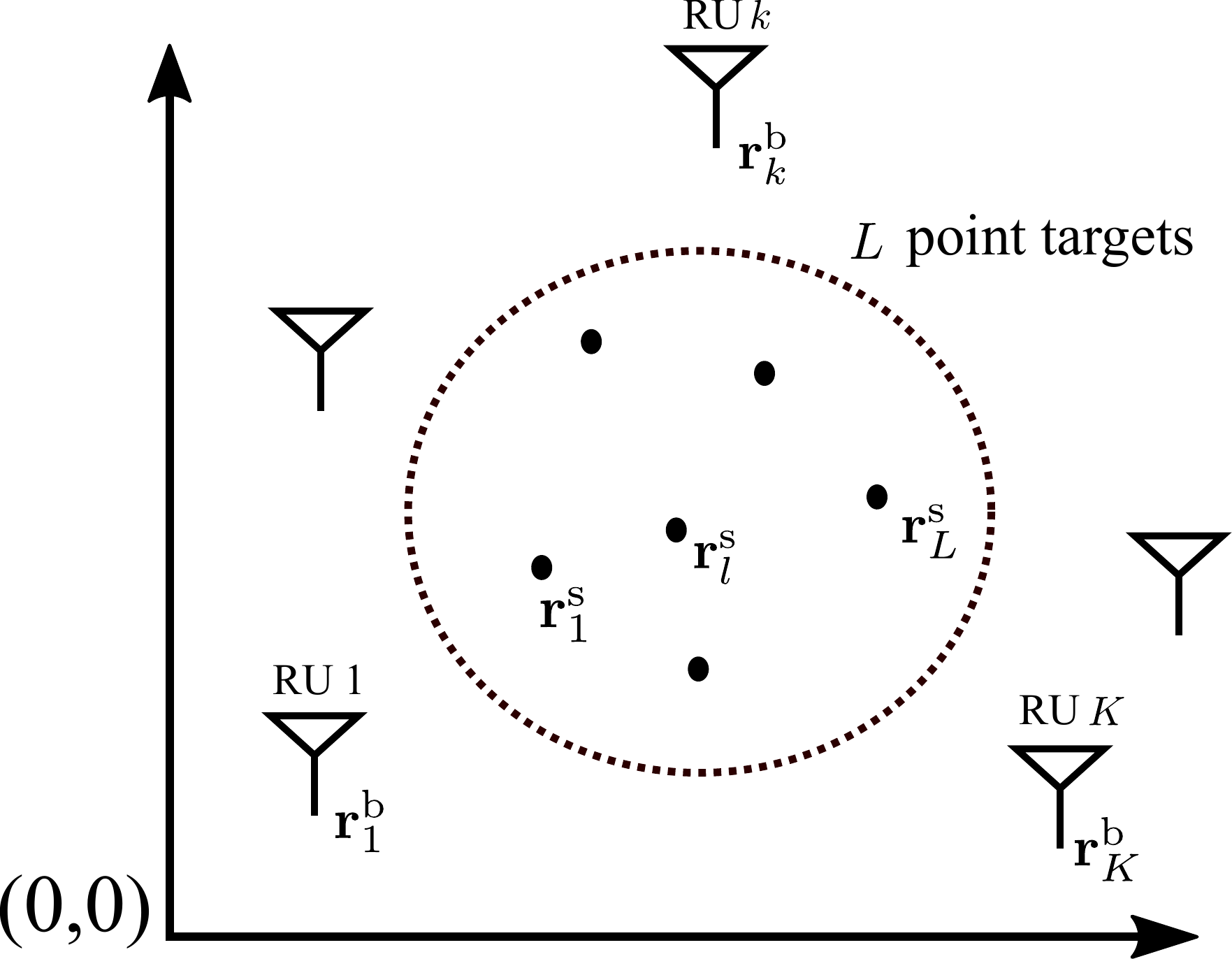}
        \caption{$K$ RUs in a cell-free network alternatively serve as illuminators and receivers to cooperatively sense $L$ point targets in a global 2D coordinate.}
        \label{fig:system_model}
        % \vspace{-5mm}
\end{figure}

\section{System model}
We consider a cell-free communication network with $K$ static RU and $L$ static radar targets to be sensed, whose locations are respectively given as two dimensional (2D) vectors $\mathbb{r}^{\text{b}}=\{\rv_k^{\rm b}|  k \in [K]\}$ and $\mathbb{r}^{\text{s}}=\{\rv^{\rm s}_l|  l \in [L]\}$ according to a global 2D coordinate, as shown in Fig.~\ref{fig:system_model}. We assume that the $k$-th RU is equipped with $M_k$ antennas in a fully digital uniform linear array (ULA). We further assume the wideband system running with OFDM signaling, where the whole bandwidth is divided into $N$ subcarriers. 
\subsection{Channel model}
Considering $S$ OFDM symbols for sensing in total, the frequency domain sensing channel model from the $i$-th RU to $k$-th RU at $s$-th time slot and $n$-th subcarrier is given as
\begin{align}
    \Hm_{i,k}[s,n]  &= \sum^L_{l=1} \rho_{l} \sqrt{D_{i,k,l}} \av_k(\theta_{k,l}) \bv_i^\herm(\phi_{i,l}) e^{-j2\pi \tau_{i,k,l} (n-1)\Delta f},\notag\\ 
    &\quad\qquad i,k \in [K], s\in [S], n\in[N],
    \label {channel_model}
\end{align}
where $\tau_{i,k,l} = \frac{d_{i,l}+d_{k,l}}{c_0}$ is the delay from $i$-th RU to $k$-th RU reflected by the $l$-th target with $c_0$ being the speed of light, and where $d_{k,l} = \|\rv^{\rm b}_k-\rv^{\rm s}_l\|_2, \forall k \in [K],l\in [L]$ being the distance between the $k$-th RU and the $l$-th target, and where $\Delta f$ is the subcarrier spacing. Moreover, we assume that all the signals are in the far field and thus the $m$-th element of steering vectors in~\eqref{channel_model} for the transmitter and receivers are respectively given as
\begin{align}
    [\av_k(\theta_{k,l})]_m &= e^{j\pi (m-1) \sin(\theta_{k,l})}, \quad \forall m \in [M_k], \\ 
    [\bv_i(\phi_{i,l})]_m &= e^{j\pi (m-1) \sin(\phi_{i,l})}, \quad \forall m \in [M_i],
\end{align}
where the antenna spacing is equal to half of carrier wavelenth $\lambda$, and where $\theta_{k,l}$ and $\phi_{i,l}$ are the angle of departure (AoD) and angle of arrival (AoA) between the $k$-th ($i$-th) RU and the $l$-th target, respectively. Furthermore,
$D_{i,k,l}$ is the effective pathloss from the $i$-th RU to the $k$-th RU through the $l$-th target, which is given as \cite{richards2005fundamentals}
\begin{equation}
    D_{i,k,l} = \frac{\lambda^2}{(4\pi)^3 d_{i,l}^2d_{k,l}^2}.
\end{equation}
Finally, $\rho_{l}$ in \eqref{channel_model} is the complex channel fading coefficient that follows a zero-mean complex Gaussian distribution as $\rho_{l} \sim \mathcal{CN}(0, \gamma_{l})$, where $\gamma_{l}$ is the RCS of the $l$-th target \cite{richards2005fundamentals}.\footnote{In this work, we assume that the RCS is isotropic and thus independent of the angle of illumination. The sensing task under non-isotropic RCS is left for future work.}
% and where we define $D_{i,k,l} \triangleq  \frac{\lambda^2}{(4\pi)^3 d_{i,l}^2d_{k,l}^2}$. Note that we assume that $\rho_{i,k,l}$ is constant over all $S$ time slots and $N$ subcarriers. We can write $\rho_{i,k,l} = \sqrt{D_{i,k,l}}\delta_l$, where $\delta_l\sim \mathcal{CN}(0, \sigma_l)$.

\subsection{Transmit signal}
For the ease of analysis, we consider the communication pilot siganls for target detection in the cell-free communication network.
We consider a beamforming (BF) pattern of the RU $k\in[K]$ with $Z_k < M_k$ BF vectors $\Fm_k=[\fv_{k,1} \dots, \fv_{k,Z_k}]$, where the BF vectors are almost orthonormal, i.e., $\Fm_k^\herm \Fm_k \approx \Id$, and where each angle is covered by a single beam, i.e.,
\begin{equation}
    |\fv_{k,i}^\herm \bv_k(\phi)| \gg 1 \implies |\fv_{k,j}^\herm \bv_k(\phi)| \approx 0, \;\forall i \neq j \in [Z_k].
\end{equation}
Then, the transmit signal from the $k$-th RU at time slot $s\in[S]$ and subcarrier $n\in[N]$ is the weighted sum of all BF vectors:
\begin{align}
    \xv_k[s,n] = \sum^{Z_k}_{z=1} w_{k,z}[s] \fv_{k,z} 
    = \Fm_k \wv_k[s],
\end{align}
where $\wv_k[s] \triangleq [w_{k,1}[s], \dots, w_{k,Z_k}[s]]^\transp$ is the vector of all weights $0\leq w_{k,z}[s]\leq 1$ that are the same for all subcarriers but can be adjusted in different time slots for different beam patterns. Given the total transmit power of the $k$-th RU as $P_k$ and assuming equal power allocation over all subcarriers, the transmit power constraint is given as 
\begin{equation}
    \sum^{Z_k}_{z=1} |w_{k,z}[s]|^2 \leq \frac{P_k}{N}, \quad \forall k\in [K], s\in[S].
\end{equation}

\subsection{Received signal}
Given the transmitted signal of the $i$-th RU at time slot $s$ and subcarrier $n$ as $\xv_i[s,n]$, the received \textbf{noiseless} signal at the $k$-th RU from the $i$-th RU is given as
% \begin{align}
%     \yv_{i,k}[s,n] &= \Hm_{i,k}[s,n]\xv_i[s,n] \\
%         &= \sum^L_{l=1} \rho_{l} \sqrt{D_{i,k,l}} \av_k(\theta_{k,l}) \underbrace{\bv_i^\herm(\phi_{i,l}) \xv_i[s,n]}_{\triangleq b_{i,l}[s,n]}\notag \\ & \qquad
%         e^{-j2\pi \tau_{i,k,l} (n-1)\Delta f} \\
%         &=\sum^L_{l=1} \rho_{l} \sqrt{D_{i,k,l}} \av_k(\theta_{k,l}) b_{i,l}[s,n] e^{-j2\pi \tau_{i,k,l} (n-1)\Delta f},
% \end{align}
\begin{align}
    &\yv_{i,k}[s,n] = \Hm_{i,k}[s,n]\xv_i[s,n] \\
        &= \sum^L_{l=1} \rho_{l} \sqrt{D_{i,k,l}} \av_k(\theta_{k,l}) \underbrace{\bv_i^\herm(\phi_{i,l}) \xv_i[s,n]}_{\triangleq b_{i,l}[s,n]}
        e^{-j2\pi \tau_{i,k,l} (n-1)\Delta f},\notag\\
        &=\sum^L_{l=1} \rho_{l} \sqrt{D_{i,k,l}} \av_k(\theta_{k,l})  b_{i,l}[s,n]
        e^{-j2\pi \tau_{i,k,l} (n-1)\Delta f},
        \label {receive_symbols}
\end{align}
% where we neglect the noise term for simplicity.
where $b_{i,l}[s,n]$ is the transmit beamforming gain between the $i$-th RU and $l$-th target at time slot $s$ and subcarrier $n$. We denote the generalized delay response vector as $\tv(\tau_{i,k,l}, b_{i,l}[s,n])$ containing the transmit beamforming gain, whose $n$-th element is given as
\begin{equation}
    [\tv(\tau_{i,k,l},  b_{i,l}[s,n])]_n =  b_{i,l}[s,n]e^{-j2\pi\tau_{i,k,l}(n-1)\Delta f}, \; \forall n \in [N]. 
\end{equation} 
Then, stacking the received noiseless signal of all $N$ subcarriers, we have 
\begin{align}
\bar{\yv}_{i,k}[s] &\triangleq \left[\yv_{i,k}[s,1]^\transp,\dots,\yv_{i,k}[s,N]^\transp\right]^\transp, \\ 
    &=\sum^L_{l=1} \rho_{l}\underbrace{\sqrt{D_{i,k,l}}  \av_k(\theta_{k,l})\otimes\tv(\tau_{i,k,l},b_{i,l}[s,n])}_{\triangleq\psiv_{i,k,l}[s]}, \\
    &=\Psim_{i,k}[s] \rhov,
\end{align}
where $\Psim_{i,k}[s] \triangleq 
[\psiv_{i,k,1}[s], \dots,  \psiv_{i,k,L}[s]]$ and $\rhov\triangleq[\rho_1,\dots,\rho_L]^\transp$ is a vector of channel fading coefficient related to the RCS.

\subsection{Illumination-Receiver scheme}
We denote the index set of RU as $\mathcal{K}\triangleq[K]$. To avoid the requirement of full-duplex capability (which may result in intractable residual self-interference that is not neglectable), we consider an alternate illumination-receiving scheme, where at each time slot $s$, the index set of RU that only illuminates (transmit signal) is denoted as $\mathcal{K}_s^{\text{t}} \subset \mathcal{K}$ and the index set of RU that only receive sensing signals is denoted as $\mathcal{K}_s^{\text{r}}= \mathcal{K} \setminus \mathcal{K}_s^{\text{t}}$. Then, the received \textbf{noisy} signal at the $k$-th RU and time slot $s$ is the sum of signals from all illuminating RU and the noise:
\begin{equation}
    \widetilde{\yv}_{k}[s] = \sum_{i\in\mathcal{K}_s^{\text{t}}} \bar{\yv}_{i,k}[s] + \nv_k[s], \quad  k\in \mathcal{K}_s^{\text{r}},
\end{equation}
where $\nv_k[s] \sim \mathcal{CN}(\mathbf{0},N_0\Id_{NM_k})$ is the additive white Gaussian noise (AWGN) with noise power $N_0$.

Since we focus on the strong cooperative sensing, we assume that a central unit (CU) can obtain the observations of all RUs and process globally. Stacking all noisy observations of all $S$ time slots, we have
\begin{equation}\label{eq:received_signal_all}
    \mathbb{y} \triangleq \begin{bmatrix}
        \widetilde{\yv}_{[\mathcal{K}_1^{\text{r}}]_1}[1] \\
        \vdots \\
        \widetilde{\yv}_{[\mathcal{K}_1^{\text{r}}]_{K^{\text{r}}_1}}[1] \\
        \widetilde{\yv}_{[\mathcal{K}_2^{\text{r}}]_{1}}[2]\\
        \vdots\\
        \widetilde{\yv}_{[\mathcal{K}_S^{\text{r}}]_{K^{\text{r}}_S}}[S]
    \end{bmatrix} = \begin{bmatrix}
        \sum_{i\in\mathcal{K}^{\text{t}}_1}\Psim_{i,[\mathcal{K}^{\text{r}}_1]_1}[1] \\
        \vdots \\
        \sum_{i\in\mathcal{K}^{\text{t}}_1}\Psim_{i,[\mathcal{K}^{\text{r}}_1]_{K^{\text{r}}_1}}[1]\\
        \sum_{i\in\mathcal{K}^{\text{t}}_2}\Psim_{i,[\mathcal{K}^{\text{r}}_2]_1}[2]\\
        \vdots \\
        \sum_{i\in\mathcal{K}^{\text{t}}_S}\Psim_{i,[\mathcal{K}^{\text{r}}_S]_{K^{\text{r}}_S}}[S]
    \end{bmatrix} \rhov + \nv,
\end{equation}
where $K^{\text{t}}_s = |\mathcal{K}^{\text{t}}_s|$, $K^{\text{r}}_s = |\mathcal{K}^{\text{r}}_s|$ and $[\mathcal{K}]_i$ denotes the $i$-th element of set $\mathcal{K}$.

\textbf{Task:} Given all received observations of $S$ time slots $\mathbb{y}$ at the CU, how to estimate the locations of $L$ target $ \mathbb{r}^{\rm s}$?

\section{Parametric fitting based location estimation}
% In order to solve the sensing problem cooperatively, we must find the unique information of targets that is independent of the RU. {\color{red} I do not understand this part ... Traditionally focused variables}, e.g., angle or delay are not such unique information, since they depend on the illuminator and receiver. It is noticed that the locations of targets in the given global 2D coordinate system are unique and independent of RU. 

In contrast to the weak cooperative sensing scheme, where each RU focuses on its own traditional variables, e.g., angle or delay (distance), and a subsequent data fusion is applied, we propose a strong cooperative sensing approach that exploits the common global 2D coordinate system of the network. 
It is noticed that the angle and delay (distance) from a given position in the network to any RU are directly obtainable based on the coordinates of that position and RU. Thus, we propose a parametric representation of the observations with a 2D grid-based location domain. The location and RCS of targets are both detected and estimated based on the grid points by fitting the parametric representation to the received signal at the CU.

We discretize the area to be sensed with very dense $Q\gg L$ grid points, whose coordinates are given as $\mathbb{r}^{\text{g}} = \{\rv^{\text{g}}_q|q\in[Q]\}$. Then, we obtain grid point based parameters $\theta_{k,q}(\rv^{\text{g}}_q), \phi_{i,q}(\rv^{\text{g}}_q), \tau_{i,k,q}(\rv^{\text{g}}_q), D_{i,k,q}(\rv^{\text{g}}_q), \; i,k \in [K], q \in [Q]$. Thus, we can approximate the received signal in \eqref{eq:received_signal_all} as a parametric representation. Concretely, denoting $\widetilde{\psiv}_{i,k,q}[s] \triangleq \sqrt{D_{i,k,q}(\rv^{\text{g}}_q)}  \av_k\left(\theta_{k,q}(\rv^{\text{g}}_q)\right) \otimes\tv\left(\tau_{i,k,q}(\rv^{\text{g}}_q),b_{i,q}(\rv^{\text{g}}_q)[s,n]\right)$ and  $\widetilde{\Psim}_{i,k}[s] \triangleq 
\left[\widetilde{\psiv}_{i,k,1}[s], \dots,  \widetilde{\psiv}_{i,k,Q}[s]\right]$
% \begin{align}
%    \widetilde{\psiv}_{i,k,q}[s] &\triangleq \sqrt{D_{i,k,q}(\rv^{\text{g}}_q)}  \av_k\left(\theta_{k,q}(\rv^{\text{g}}_q)\right) \notag \\ 
%    &\qquad\otimes\tv\left(\tau_{i,k,q}(\rv^{\text{g}}_q),b_{i,q}(\rv^{\text{g}}_q)[s,n]\right), \\
%    \widetilde{\Psim}_{i,k}[s] &\triangleq 
% \left[\widetilde{\psiv}_{i,k,1}[s], \dots,  \widetilde{\psiv}_{i,k,Q}[s]\right],
% \end{align}
we have 
\begin{align}
\mathbb{y} &\approx
\underbrace{
\begin{bmatrix}
        \sum_{i\in\mathcal{K}^{\text{t}}_1}\widetilde{\Psim}_{i,[\mathcal{K}^{\text{r}}_1]_1}[1] \\
        \vdots \\
        \sum_{i\in\mathcal{K}^{\text{t}}_1}\widetilde{\Psim}_{i,[\mathcal{K}^{\text{r}}_1]_{K^{\text{r}}_1}}[1]\\
        \sum_{i\in\mathcal{K}^{\text{t}}_2}\widetilde{\Psim}_{i,[\mathcal{K}^{\text{r}}_2]_1}[2]\\
        \vdots \\
        \sum_{i\in\mathcal{K}^{\text{t}}_S}\widetilde{\Psim}_{i,[\mathcal{K}^{\text{r}}_S]_{K^{\text{r}}_S}}[S]
    \end{bmatrix}}_{\triangleq\Am} \widetilde{\rhov} + \nv=\Am\widetilde{\rhov}+\nv, \label{eq:standard_form}
\end{align} 
where $\Am$ is the grid-based sensing matrix and $\widetilde{\rhov}\in\bC^Q$ is a sparse vector with sparsity $L$ indicating the fading coefficients of all grid points. Note that the approximation in \eqref{eq:standard_form} is due to the fact that in practice very likely we have a mismatched model, where targets are not located on the grid.  
% {\color{red} why we have this approximation?}

Given the standard form in \eqref{eq:standard_form}, the task is then to estimate the support (for location detection) and corresponding values (for RCS-related fading estimation) of the non-zero elements in $\widetilde{\rhov}$. Knowing $\widetilde{\rhov}$ is sparse,
we can either treat $\widetilde{\rhov}$ as deterministic unknown quantities and apply compressed sensing (CS) methods, or as random variables that follow some prior and apply sparse Bayesian learning (SBL).

\subsection{CS-based solution}
Based on the sparse parametric form in   \eqref{eq:standard_form}, we can formulate a CS problem by fitting the grid-based representation to the noisy samples. An example is the LASSO problem:
\begin{align}
    \widetilde{\rhov}^\star = \argmin_{\widetilde{\rhov}\in \bC^{Q}} \quad \|\Am\widetilde{\rhov} - \mathbb{y}\|^2_2 + \lambda \|\widetilde{\rhov}\|_1,
\end{align}
where $\lambda$ is the LASSO regularizer, which can be solved using a standard LASSO solver. However, the optimal choice of $\lambda$ is usually unknown. In our simulation, we will use the well-known greedy CS algorithm orthogonal matching pursuit (OMP) \cite{tropp2007signal} as a representative of the CS-based solution.

\subsection{SBL-based solution}
By treating $\widetilde{\rhov}_k$ as a random vector variable, we assume that it follows a Gaussian prior $p(\widetilde{\rhov}; \Gammam)$
\begin{align}\label{eq:Gaussian_prior}
    p(\widetilde{\rhov}; \Gammam) = \prod^Q_{i=1}\frac{1}{\pi \gamma_{i}} e^{-\frac{|\widetilde{\rho}_{i}|^2}{\gamma_{i}}},
\end{align}
where the hyperparameter $\gamma_{i} \geq 0$ denotes the prior variance of $\widetilde{\rho}_{i} \triangleq [\widetilde{\rhov}]_i$ and $\Gammam =  \diag(\gammav) \in \bR^{Q\times Q}$ with $\gammav \triangleq [\gamma_{1}, \dots, \gamma_{Q}]^\transp$. The SBL solution is to find the prior $p(\widetilde{\rhov}; \Gammam)$ that maximizes the Bayesian evidence $p(\mathbb{y};\Gammam)$, which leads to the location estimation. Concretely, given $p(\widetilde{\rhov}; \Gammam)$ in \eqref{eq:Gaussian_prior}, the minus log-likelihood objective for maximum-likelihood (ML) estimation of the hyperparameter vector $\gammav$ is given as 
\begin{align} \label{eq:minus_log_likelihood}
    -\log p(\mathbb{y};\Gammam) = \log(\det(\Sigmam_{\mathbb{y}})) + \mathbb{y}^\herm \Sigmam_{\mathbb{y}}^{-1}\mathbb{y},
\end{align}
where the constant term is ignored, and $\Sigmam_{\mathbb{y}}$ is the covariance matrix of  $\mathbb{y}$. It can be shown that the objective function in \eqref{eq:minus_log_likelihood} has a convex-plus-concave structure with respect to $\gammav$, which can be solved using majorization-minimization (MM) methods \cite{sun2016majorization}. Among many MM methods, we use the expectation maximization (EM) algorithm \cite{sun2016majorization} to iteratively solve it with the advantage of relatively low complexity.

Let $\widehat{\Gammam}^{(\ell)}$ denote the estimated $\Gammam$ in the $\ell$-th iteration. The expectation (E-step) in the $\ell$-th iteration evaluates the average log-likelihood of the complete data set $\{\mathbb{y},\widetilde{\rhov}\}$, given as 
\begin{align}
    \mathcal{L}\left(\Gammam|\widehat{\Gammam}^{(\ell)}\right) & = \bE_{\widetilde{\rhov}|\mathbb{y};\widehat{\Gammam}^{(\ell)}} \left[\log p(\mathbb{y},\widetilde{\rhov};\Gammam)\right] \\
    &=  \bE_{\widetilde{\rhov}|\mathbb{y};\widehat{\Gammam}^{(\ell)}} \left[\log p(\mathbb{y}|\widetilde{\rhov})+\log p(\widetilde{\rhov};\Gammam)\right], \label{eq:second_term}
\end{align}
where the first term in \eqref{eq:second_term} is independent of $\gammav$ and thus is ignored in the maximization step (M-step). To evaluate the second term, we employ the \textit{a posterior}  probability density of $\widetilde{\rhov}$, which is given as \cite{wipf2004sparse, tipping2001sparse}
\begin{align}
\text{E-step:}    \quad &p\left(\widetilde{\rhov}|\mathbb{y};\widehat{\Gammam}^{(\ell)}\right) \sim \mathcal{CN}\left(\muv_{\widetilde{\rhov}}^{(\ell)}, \Sigmam_{\widetilde{\rhov}}^{(\ell)}\right) \label{eq:posterior}\\
    \text{with} \quad  &\muv_{\widetilde{\rhov}}^{(\ell)} = \Sigmam_{\widetilde{\rhov}}^{(\ell)} \Am^\herm \mathbb{y} \in \bC^Q, \label{eq:E_step_mu}\\ &\Sigmam_{\widetilde{\rhov}}^{(\ell)} = \left(\Am^\herm \Am + \left(\widehat{\Gammam}^{(\ell)}\right)^{-1}\right)^{-1} \in \bC^{Q\times Q}.\label{eq:E_step_Sigma}
\end{align}
Using \eqref{eq:Gaussian_prior} and \eqref{eq:posterior} the second term in \eqref{eq:second_term} can be written as
\begin{align}
    \mathcal{L}_{\gammav}^{(\ell)}&\triangleq \bE_{\widetilde{\rhov}|\mathbb{y};\widehat{\Gammam}^{(\ell)}} \left[\log p(\widetilde{\rhov};\Gammam)\right] \notag \\ 
    & = \sum^Q_{i=1}\left(-\log (\pi\gamma_{i}) - \frac{\bE_{\widetilde{\rhov}|\mathbb{y};\widehat{\Gammam}^{(\ell)}}\left[|[\widetilde{\rhov}]_i|^2\right]}{\gamma_{i}}\right), \\
    &= \sum^Q_{i=1}\left(-\log (\pi\gamma_{i}) - \frac{\left|\left[\muv_{\widetilde{\rhov}}^{(\ell)}\right]_i\right|^2 + \left[\Sigmam_{\widetilde{\rhov}}^{(\ell)}\right]_{i,i} }{\gamma_{i}}\right). \label{eq:second_term_decoupled}
\end{align}
The estimated hyperparameter vector $\widehat{\gammav}^{(\ell+1)}$ is updated in the M-step by maximizing the resulting objective function in \eqref{eq:second_term_decoupled} with respect to $\gammav$, i.e.,
\begin{align}
     \widehat{\gammav}^{(\ell+1)} = \argmax_{\gammav} \; \mathcal{L}_{\gammav}^{(\ell)}.
\end{align}
It is noticed from \eqref{eq:second_term_decoupled} that the maximization is decoupled with respect to each component $\gamma_{i}$ and the optimality is obtained in closed form by setting the partial derivative to zero as
\begin{align}\label{eq:M_step_gamma}
    \text{M-step:} \quad \widehat{\gamma}_{i}^{(\ell+1)} = \left|\left[\muv_{\widetilde{\rhov}}^{(\ell)}\right]_i\right|^2 + \left[\Sigmam_{\widetilde{\rhov}}^{(\ell)}\right]_{i,i} , \; \forall i\in [Q].
\end{align}
The initial point of the hyperparameter vector can be set to all ones, i.e., $\widehat{\gammav}^{(0)} = \mathbf{1}$. The E-step and M-step are alternately applied until the predefined maximum number of iterations $G$ is reached, or the stop condition $\|\widehat{\gammav}^{(\ell+1)} - \widehat{\gammav}^{(\ell)}\| /\|\widehat{\gammav}^{(\ell)}\| < \epsilon$ is met, where $\epsilon > 0$ is the predefined stop condition.
When the EM algorithm is converged, the target locations are estimated by finding $L$ dominant maximizers of the resulting $\widehat{\gammav}^{(\ell)}$ as:
\begin{align}\label{eq:maxk}
    \widehat{\mathbb{r}}^{\text{s}}  = \{\rv^{\text{g}}_i| i\in\mathcal{I}_L\}\subseteq \mathbb{r}^{\text{g}}, \;
    \text{with}\;\; \mathcal{I}_L  = \arg\; \underset{_{i\in[Q]}}{\mathtt{maxL}} \; \widehat{\gammav}^{(\ell)},
\end{align}
where $\mathtt{maxL}$ gives the $L$ largest elements of a vector.  
In the case without the prior knowledge of the number of targets $L$, we apply constant false alarm rate (CFAR) detection \cite{richards2005fundamentals} to jointly detect the number and locations of targets with a low false alarm rate (FAR), which is given as
\begin{align}\label{eq:cfar}
     \widehat{\mathbb{r}}^{\text{s}} = \{\widehat{\rv}^{\text{s}}_1,\dots,\widehat{\rv}^{\text{s}}_{\widehat{L}}\} = \text{cfar}\left(\widehat{\gammav}^{(\ell)}\right).
\end{align}
The complete SBL-based approach is given in Algorithm~\ref{alg:SBL-EM}.

\vspace{-3mm}
\begin{algorithm}[H]
\caption{SBL-EM for location estimation}\label{alg:SBL-EM}
\begin{algorithmic}[1] 
\State{Initialize $\widehat{\gammav}^{(0)} = \mathbf{1}$}
\For{$\ell=0,\dots,G$} 
    \State{\textbf{E-step:} update posterior covariance $\Sigmam_{\widetilde{\rhov}}^{(\ell)} \gets$ \eqref{eq:E_step_Sigma}} 
    \State{\textbf{E-step:} update posterior mean $\muv_{\widetilde{\rhov}}^{(\ell)} \gets$ \eqref{eq:E_step_mu}}
    \State{\textbf{M-step:} update hyperparameters $\widehat{\gammav}^{(\ell+1)} \gets$ \eqref{eq:M_step_gamma}}
    \If{$\|\widehat{\gammav}^{(\ell+1)} - \widehat{\gammav}^{(\ell)}\|/{\|\widehat{\gammav}^{(\ell)}\|} < \epsilon$} 
        \State {Break for-loop}
    \EndIf
\EndFor
\If{ The number of target $L$ is known}
    \State{Obtain the estimated target locations $\widehat{\mathbb{r}}^{\text{s}} \gets$ \eqref{eq:maxk}}
\Else
    \State{Using CFAR to detect $\widehat{\mathbb{r}}^{\text{s}} \gets$ \eqref{eq:cfar}}
\EndIf
\State{\Return$\widehat{\mathbb{r}}^{\text{s}}$}
\end{algorithmic}
\end{algorithm}
\vspace{-5mm}

\section{Theoretical analysis with PEP}
In this section, we provide a theoretical analysis on the estimation error of the proposed grid-based approach using PEP under a matched model (with only on-grid targets)\footnote{Note that the estimation error due to the mismatch of the grid points and true locations is not considered in this PEP-based analysis. The unconsidered mismatch error should be negligible under a sufficient dense grid.}. 

Supposing $L$ targets are exactly located on the grid with $Q$ grid points, i.e., the fading coefficient vector $\widetilde{\rhov} \in \bC^Q \sim \mathcal{CN}(\mathbf{0}, \Cm)$ have exactly $L$ non-zero elements, where $\Cm$ is a diagonal matrix containing the RCS, we can rewrite the fading coefficients as $\widetilde{\rhov} = \diag(\widetilde{\rhov})\qv =\diag(\qv)\widetilde{\rhov}$, where $\qv \in \{0,1\}^Q$ is a binary vector indicating the active grid points. Then, we have $\mathbb{y} = \Am \diag(\qv)\widetilde{\rhov} + \nv$, and the probability of erroneously estimated $\widehat{\qv}$ from the true sequence of locations $\qv$ is upper-bounded by \cite{tarokh1998space}
\begin{align}\label{eq:PEP}
    \text{Pr}(\qv \rightarrow \widehat{\qv} | \widetilde{\rhov}) \leq e^{-\frac{d^2(\qv,\widehat{\qv})}{4N_0}},
\end{align}
where $d^2(\qv,\widehat{\qv})$ is the conditional Euclidean distance between two sequences, which is given as \cite{tarokh1998space, li2021performance}
\begin{align}
   \hspace{-3mm} d^2(\qv,\widehat{\qv}) &= \|\Am\diag(\ev)\widetilde{\rhov}\|^2, \\
    &=\widetilde{\rhov}^\herm \diag(\ev)\Am^\herm\Am\diag(\ev)\widetilde{\rhov}, \\
    &= \widetilde{\rhov}^\herm \Cm^{-\frac{1}{2}}\underbrace{\Cm^{\frac{1}{2}} \diag(\ev)\Am^\herm\Am\diag(\ev)\Cm^{\frac{1}{2}}}_{\triangleq \widetilde{\Am}(\ev)}\underbrace{\Cm^{-\frac{1}{2}}\widetilde{\rhov}}_{\triangleq \gv}, \label{eq:Euc_distance_conditional}
\end{align}
where $\ev \triangleq \qv - \widehat{\qv}$ is the error sequence, $\widetilde{\Am}(\ev)$ is a Hermitian positive semi-definite matrix, and $\gv \sim \mathcal{CN}(\mathbf{0}, \mathbf{I})$ is the normalized fading vector. 
Substituting \eqref{eq:Euc_distance_conditional} into \eqref{eq:PEP}, we obtain the upper bound to the conditional PEP (CPEP): 
\begin{align}\label{eq:PEP_conditional}
    \text{Pr}(\qv \rightarrow \widehat{\qv} | \gv) \leq e^{-\frac{\gv^\herm \widetilde{\Am}(\ev)\gv}{4N_0}}.
\end{align}

Now, we derive the upper bound to unconditional PEP (UPEP) by averaging over the channel fading coefficients $\gv$. 
We denote the eigenvalue decomposition of $\widetilde{\Am}(\ev)$ as $\widetilde{\Am}(\ev) = \Um \Lambdam \Um^\herm$, where $\Lambdam = \diag(\lambdav)$ contains $Q$ eigenvalues $\lambdav=[\lambda_1, \dots, \lambda_Q]^\transp$ in descending order of $\widetilde{\Am}(\ev)$ and $\Um = [\uv_1, \dots, \uv_Q]$ contains $Q$ corresponding eigenvectors. Supposing that the rank of $\widetilde{\Am}(\ev)$ is $r = \mathtt{rank}(\widetilde{\Am}(\ev))\leq Q$, we have $\gamma_i > 0$ if $i \in [r]$ and $\gamma_i = 0$ otherwise. Then, the term $\gv^\herm \widetilde{\Am}(\ev)\gv$ in \eqref{eq:PEP_conditional} can be reformulated as 
\begin{align}
    \gv^\herm \widetilde{\Am}(\ev)\gv &= \gv^\herm \left(\sum^{r}_{i=1} \lambda_i \uv_i \uv_i^\herm \right) \gv, \\
    &= \sum^{r}_{i=1} \lambda_i \gv^\herm \uv_i \underbrace{\uv_i^\herm \gv}_{\triangleq \widetilde{g}_i}= \sum^{r}_{i=1} \lambda_i |\widetilde{g}_i|^2,
\end{align}
where $\widetilde{\gv} = [\widetilde{g}_1, \dots, \widetilde{g}_r]^\transp \sim \mathcal{CN}(\mathbf{0}, \mathbf{I})$ are independent complex Gaussian random variables with covariance $\bE[\widetilde{\gv} \widetilde{\gv}^\herm] = \Um^\herm \Cm \Um = \mathbf{I}$. Since $\{|\widetilde{g}_i|\}_{i=1}^Q$ are independent mean-zero variables, they follow independent Rayleigh distributions with probability density function (PDF) $p(|\widetilde{g}_i|) = 2|\widetilde{g}_i| e^{-|\widetilde{g}_i|^2  }, i \in [Q]$. 
Given the PDF of $|\widetilde{g}_i|$ and CPEP in \eqref{eq:PEP_conditional}, we can obtain the UPEP by averaging over $|\widetilde{g}_i|$ as
\begin{align}
    \text{Pr}(\qv\rightarrow\widehat{\qv}) &= \int \text{Pr}(\qv \rightarrow \widehat{\qv}|\gv)p(\gv)\text{d}\gv,\\
    &\leq\prod_{i=1}^r\int_0^{\infty} e^{-\frac{\lambda_i|\widetilde{g}_i|^2}{4N_0}} p(|\widetilde{g}_i|)\text{d}|\widetilde{g}_i|, \label{eq:PEP_independent}\\
    &=\prod^r_{i=1} \left(1+\frac{\lambda_i}{4N_0}\right)^{-1},\label{eq:PEP_unconditional}
    % &\leq \prod^r_{i=1} \left(\frac{\lambda_i}{4N_0}\right)^{-1},\\
    % &=\left(\frac{1}{4N_0}\right)^{-r} \left(\prod^r_{i=1}\lambda_i\right)^{-1},
\end{align}
which shows that the error probability decreases with a higher SNR and a larger effective determinant (product of all non-zero eigenvalues) of the matrix $\widetilde{\Am}(\ev)$. In the simulation, we will consider the union bound of UPEP by summing up all UPEP of possible $\ev$, given a sensing matrix $\Am$. Note that the union bound of UPEP could be larger than one, and it is normally far from the true detection performance~\cite{tarokh1998space, Stefanov2003performance}. However, they are very useful for providing a qualitative comparison \cite{tarokh1998space, li2021performance}.   

\section{Numerical results}
In the simulation, we consider a cell-free communication network with $K=3$ RUs located in the vertices of an equilateral triangle with the global coordination $(0,0)$, $(100,0)$, and $(50,86)$. All RUs are equipped with $M_k = 16, k\in[K]$ antennas in ULA, whose normal lines pass through the center of the equilateral triangle. The carrier frequency is $10$ GHz with $N=16$ subcarriers and a total bandwidth of $160$ MHz. The number of target $L$ is a random integer from $3$ to $7$. All targets are uniformly located in a square with diagonal vertices at $(25,20)$ and $(75,70)$, and their RCS are set to $\gamma_l=20$ dBm, $l\in[L]$. Unless otherwise specified, the grid size is set to $Q=20\times20$. Please see Fig.~\ref{fig:point_targets} for topology illustration. The numerical results are averaged over 200 channel realizations and random beam directions.

\subsection{Illumination pattern and BF pattern}
We consider an alternating illumination-receiver scheme, where each RU in turn illuminates once while the other RUs receive, i.e., $ \mathcal{K}^{\text{t}}_s = \{s\}, s\in [S =3]$.

We set $Z_k = 10$ BF directions for each RU covering an angle range of $[-\frac{\pi}{2}, \frac{\pi}{2}]$ and compare two BF patterns, namely the equal and random power BF:  
\begin{itemize}
    \item \textbf{Equal power BF:} the power of all $Z_k$ BF directions are equally allocated as $w^2_{k,z} = \frac{P_k}{NZ_k},\; \forall z\in[Z_k]$.
    \item \textbf{Random power BF:} the power of all $Z_k$ BF directions are randomly allocated as $w^2_{k,z} = \alpha \omega_{k,z}$, where $\omega_{k,z} \sim \mathtt{Unif}(0,1)$ and $\alpha$ is set to satisfy $\sum_{z=1}^{Z_k}w^2_{k,z} = \frac{P_k}{N}$.
\end{itemize}

\subsection{Evaluation metric and result analysis}
\subsubsection{On-grid targets} 
\begin{figure*}[ht!]
\centering
        \begin{subfigure}[b]{0.31\textwidth}
        \includegraphics[width=1.05\columnwidth]{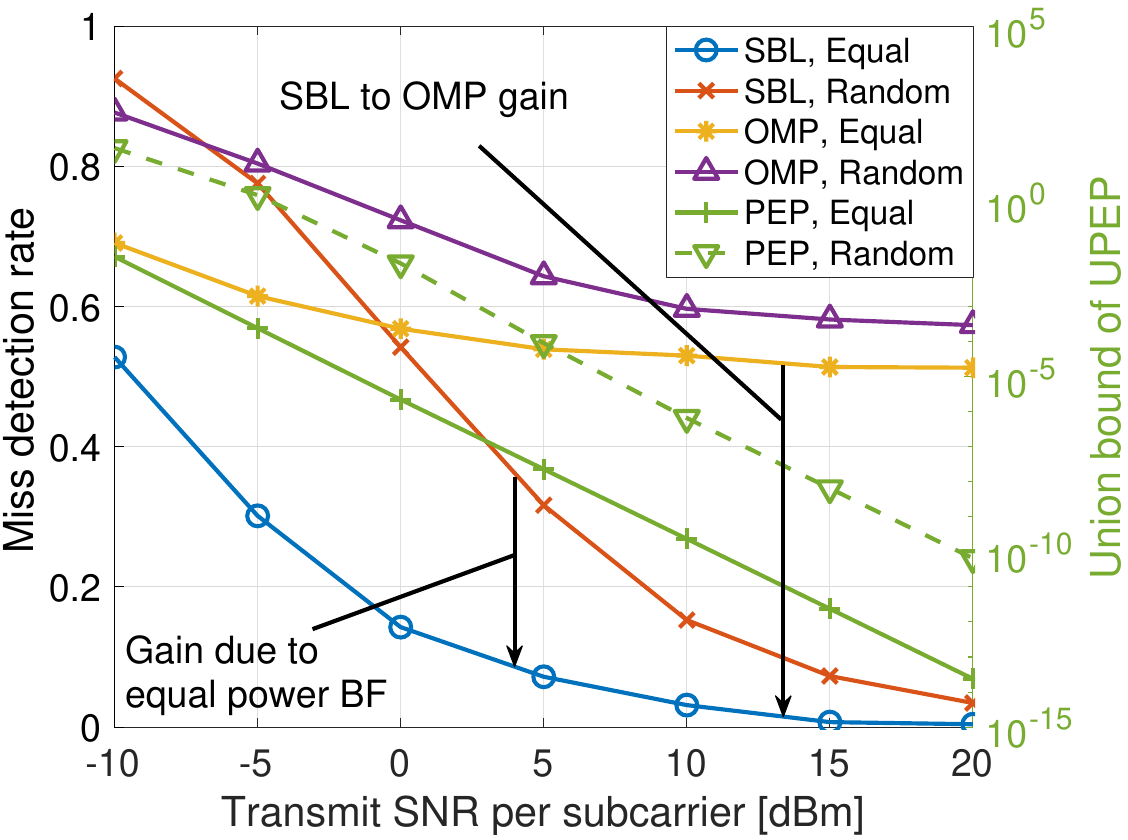}
        % \vspace{-1mm}
        \caption{MDR and PEP, on-grid, known $L$}
        \label{fig:mdr_pep}
        \end{subfigure}
        ~
        \begin{subfigure}[b]{0.31\textwidth}
        \includegraphics[width=1.05\columnwidth]{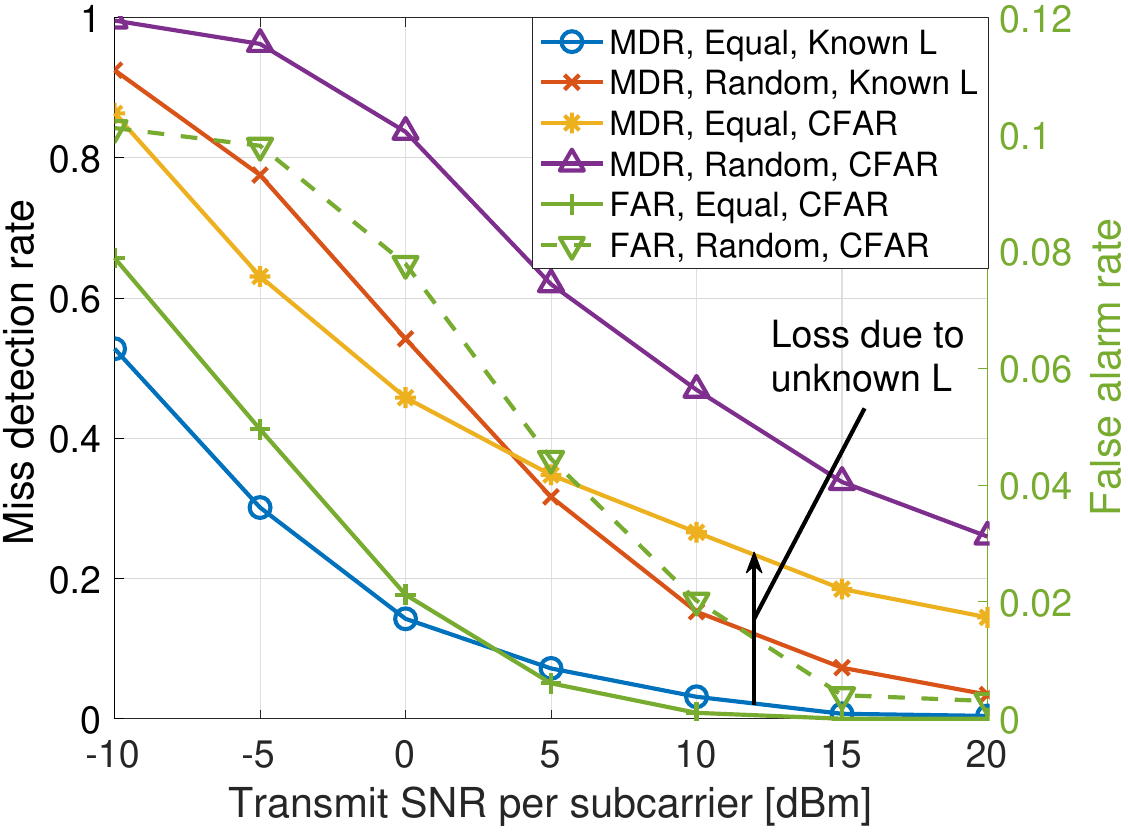}
        % \vspace{-1mm}
        \caption{MDR and FAR, only SBL, on-grid}
        \label{fig:mdr_far}
        \end{subfigure}
        ~
        \begin{subfigure}[b]{0.31\textwidth}
        \includegraphics[width=1\columnwidth]{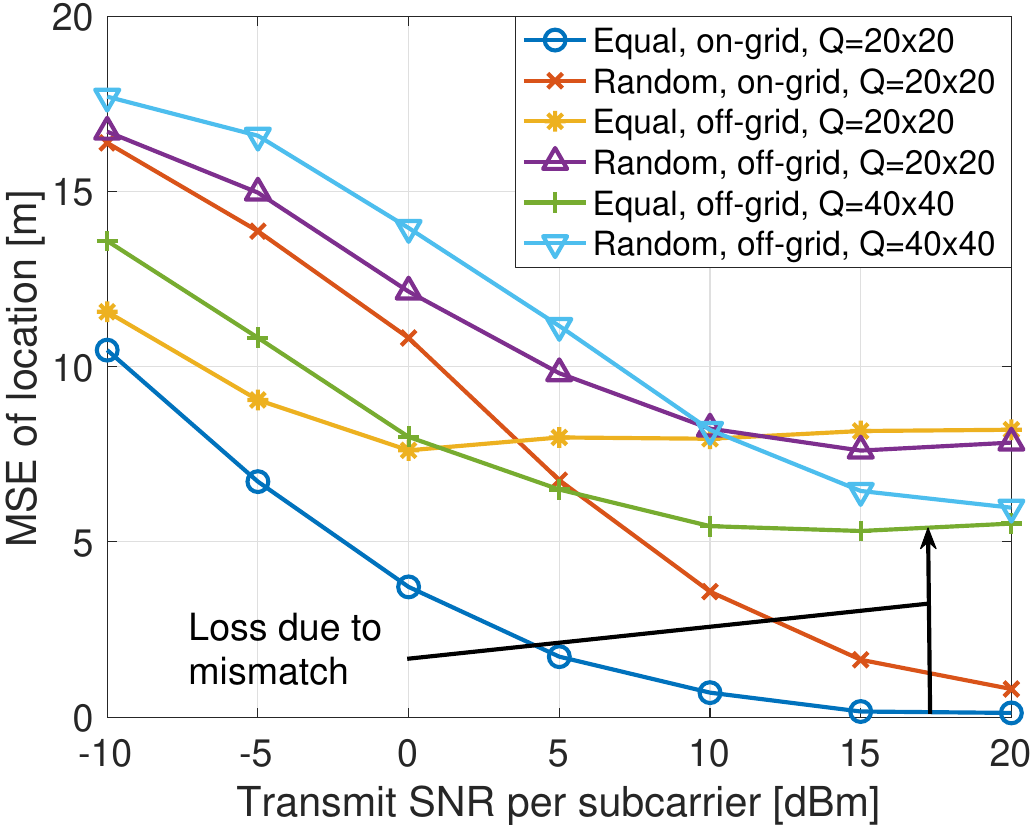}
        % \vspace{-1mm}
        \caption{MSE, only SBL, on- and off-grid }
        \label{fig:mse}
        \end{subfigure}
        % \vspace{-2mm}
 \caption{Results under various SNR: (a) compares SBL, OMP, and PEP; (b) compares CFAR; (c) compares on- and off-grid. }
     \label{fig:snr}
     % \vspace{-3mm}
\end{figure*}

For the on-grid targets, we evaluate the detection performance in terms of miss detection rate (MDR) and false alarm rate (FAR):
\begin{itemize}
    \item \textbf{MDR:} number ratio between the missed detected targets and the total targets.
    \item \textbf{FAR:} number ratio between the ``ghost'' detected targets (the detected location has no target) and the total targets.
\end{itemize}
We first show the result of on-grid targets with known $L$ in Fig.~\ref{fig:mdr_pep}.\footnote{Note that in the case with a known number of targets $L$, MDR is always equal to FAR, and therefore we only show MDR in Fig.~\ref{fig:mdr_pep}.}  We observe that under the tested illumination-receiver scheme, where \textbf{only three RU alternately illuminates only once}, the equal power BF outperforms random power BF in terms of lower MDR in all tested SNR levels both for SBL and OMP results, due to a better macro diversity. This advantage of omnidirectional BF under the tested setting is also verified by the UPEP result. The performance gap becomes smaller with the increase of SNR. Moreover, SBL presents a significant performance gain compared to OMP. In particular, the SBL-MDR of both BF patterns converges to almost zero under $20$ dBm SNR, while the OMP under equal power BF can only produce an MDR of $0.5$. Given such poor performance of OMP, we only provide results of SBL in the following evaluation results to make the figures easier to read. 

We then compare the results under known and unknown $L$ in Fig.~\ref{fig:mdr_far}, where the result of unknown $L$ is obtained by using the MATLAB function $\mathtt{cfar}$ in Phased Array System Toolbox, with the setting of $\mathtt{ProbabilityFalseAlarm}= 10^{-5}$. Although the resulting FAR under high SNR, e.g., $20$ dBm, reduces to almost zero because of the given very low false alarm probability $10^{-5}$, a performance loss of roughly $0.2$ MDR is observed due to the estimation of the number of targets using CFAR.

\subsubsection{Off-grid targets}
For the off-grid targets, we only consider the case with known $L$ and use the MSE of location as the evaluation metric, which is defined as   
\begin{align}
    \text{MSE} = \bE_{\mathbb{r}^{\text{s}}}\left[ \min_{ j\in[\Perm{L}{L}]} \;\frac{1}{L}\sum^L_{i=1} \|\rv_i^{\text{s}} -\left[\mathcal{P}_j (\widehat{\mathbb{r}}^{\text{s}})\right]_i\|_2\right],
\end{align}
where $\mathcal{P}_j(\widehat{\mathbb{r}}^{\text{s}})$ gives the $j$-th permutation of $\widehat{\mathbb{r}}^{\text{s}}$ out of $L!$ possibilities, and $\left[\mathcal{P}_j (\widehat{\mathbb{r}}^{\text{s}})\right]_i$ is the $i$-th element of $\mathcal{P}_j (\widehat{\mathbb{r}}^{\text{s}})$.

The results of MSE of locations for on-grid and off-grid targets are presented in Fig.~\ref{fig:mse}, where the results of off-grid targets based on a finer grid size of $Q=40\times40$ are also provided. We first observe a significant performance loss due to the mismatch of the grid points and target locations. Furthermore, by doubling the grid size from $Q=20\times20$ to $Q=40\times40$, the resulting MSE decreases due to reducing the distance between two neighboring grid points. Please also see Figs.~\ref{fig:off-20} and \ref{fig:off-40} as well as the explanation in the next part for an intuitive demonstration of the effect of grid size. 

\subsection{Detection examples}
\begin{figure*}[ht!]
\centering
        \begin{subfigure}[b]{0.31\textwidth}
        \includegraphics[width=1\columnwidth]{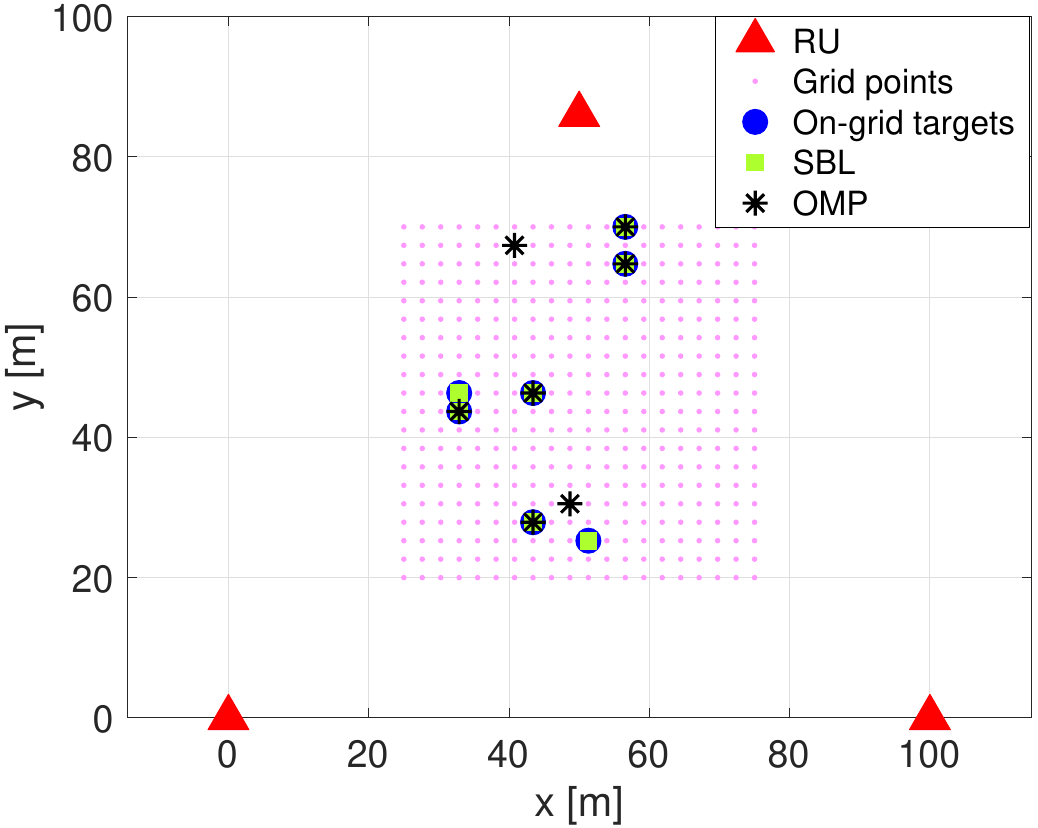}
        \caption{On-grid, $Q=20\times20$}
        \label{fig:on-grid}
        \end{subfigure}
        ~
        \begin{subfigure}[b]{0.31\textwidth}
        \includegraphics[width=1\columnwidth]{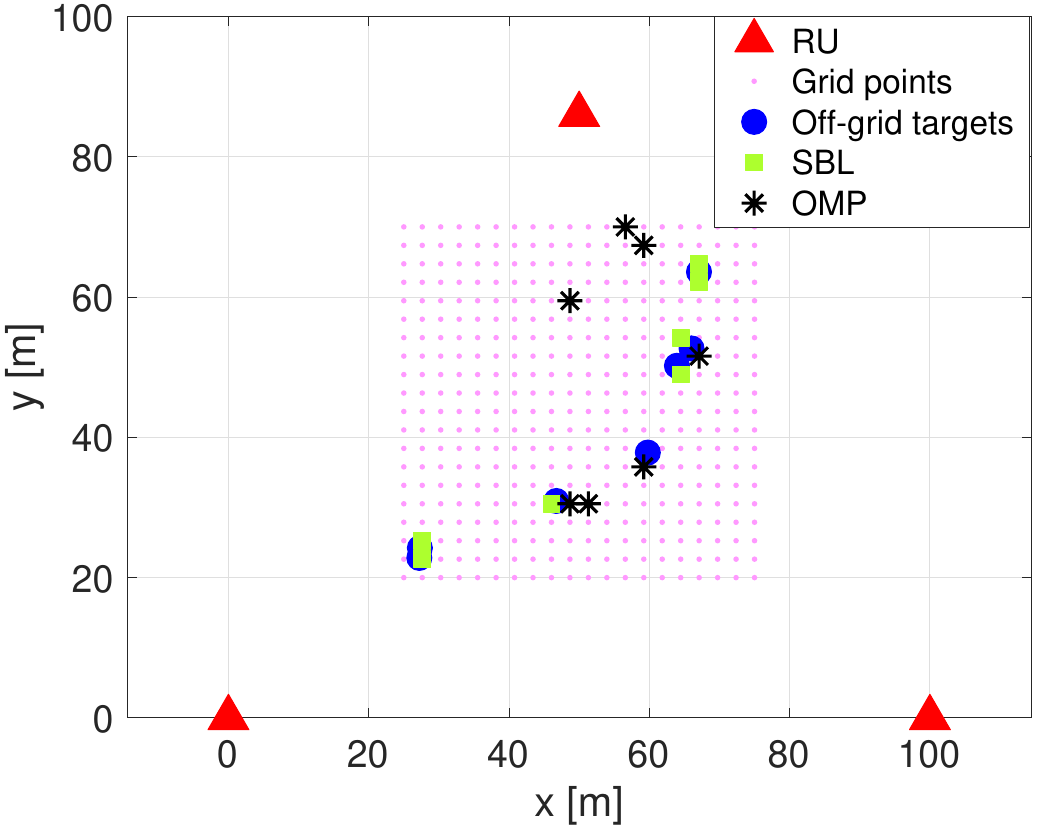}
        \caption{Off-grid, $Q=20\times20$}
        \label{fig:off-20}
        \end{subfigure}
        ~
        \begin{subfigure}[b]{0.31\textwidth}
        \includegraphics[width=1\columnwidth]{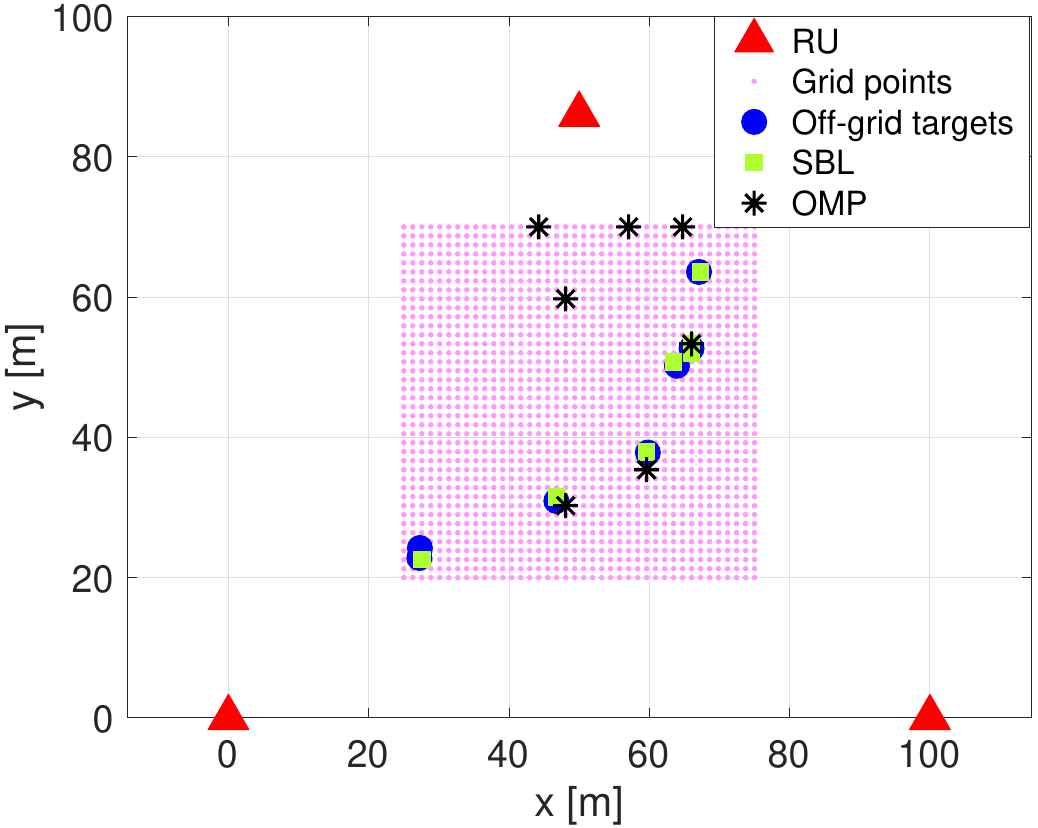}
    \caption{Off-grid, $Q=40\times40$ }
        \label{fig:off-40}
        \end{subfigure}
        % \vspace{-2mm}
 \caption{Detection examples under $20$ dBm SNR with on-grid targets in (a); and off-grid targets in (b) and (c). }
     \label{fig:point_targets}
     \vspace{-3mm}
\end{figure*}

In Fig.~\ref{fig:point_targets} we show three examples of detection under $20$ dBm SNR. In Fig.~\ref{fig:on-grid}, seven on-grid targets (blue circles) located in a $Q=20\times20$ grid (light pink dots) are all correctly detected by the proposed SBL-based algorithm (green squares), while two targets are missed detected by OMP (black stars). In Fig.~\ref{fig:off-20}, SBL misses one of seven off-grid targets, while OMP only detects three correctly and totally fails to detect the others. The same target locations and strengths are given in Fig.~\ref{fig:off-40} with double grid size, $Q=40\times40$. We observe that with a finer grid, SBL performs further better, although it still misses one target in the lower-left corner due to the proximity of two targets. In contrast, the performance of OMP does not improve with a finer grid, indicating the advantage of SBL.

\section{Conclusion}
This work focused on the strong cooperative sensing problem in distributed ISAC networks. We have formulated the cooperative multistatic target detection problem as a compressed sensing problem by exploiting the global common coordinate. The proposed SBL-based method directly detected the locations of the targets from all received signals of different illuminator-receiver pairs. We provided a PEP-based analysis of the on-grid detection error, which numerically validated the advantage of omnidirectional BF compared to random directional BF under the tested case with short-time observations. 

% performed the comparison of two different BF patterns under the tested setting that validated  
% Additionally, numerical examples also show the possibility of applying the proposed SBL-based approach for space imaging, which would be a main extension in future work.    

{\small
	\bibliographystyle{IEEEtran}
	\bibliography{references}
}

\end{document}

%% file: symbols.tex
\usepackage{mathbbol}

\def\mindex#1{\index{#1}}

\def\sq{\hbox{\rlap{$\sqcap$}$\sqcup$}}
\def\qed{\ifmmode\sq\else{\unskip\nobreak\hfil
\penalty50\hskip1em\null\nobreak\hfil\sq
\parfillskip=0pt\finalhyphendemerits=0\endgraf}\fi\medskip}

\long\def\defbox#1{\framebox[.9\hsize][c]{\parbox{.85\hsize}{%
\parindent=0pt
\baselineskip=12pt plus .1pt      % STYLE
\parskip=6pt plus 1.5pt minus 1pt % CHANGES
 #1}}}

\long\def\beginbox#1\endbox{\subsection*{}%
\hbox{\hspace{.05\hsize}\defbox{\medskip#1\bigskip}}%
\subsection*{}}

\def\endbox{}

\def\diag{{\text{diag}}}

\newsavebox{\junk}
\savebox{\junk}[1.6mm]{\hbox{$|\!|\!|$}}

\def\det{{\mathop{\rm det}}}

\def\argmin{\mathop{\rm arg\, min}}

\def\argmax{\mathop{\rm arg\, max}}

\def\bC{{\mathbb C}}

\def\bE{{\mathbb E}}

\def\bR{{\mathbb R}}

\def\bfmath#1{{\mathchoice{\mbox{\boldmath$#1$}}%
{\mbox{\boldmath$#1$}}%
{\mbox{\boldmath$\scriptstyle#1$}}%
{\mbox{\boldmath$\scriptscriptstyle#1$}}}}

\def\bfmY{\bfmath{Y}}

\def\bfmhhaY{\bfmath{\hhaY}} %\widehat{\widehat{Y}}}}
\def\bfmhhaY{\hbox to 0pt{$\widehat{\bfmY}$\hss}\widehat{\phantom{\raise 1.25pt\hbox{$\bfmY$}}}}

\def\til={{\widetilde =}}

 \def\FRAC#1#2#3{\genfrac{}{}{}{#1}{#2}{#3}}

\def\ddtp{{\mathchoice{\FRAC{1}{d^{\hbox to 2pt{\rm\tiny +\hss}}}{dt}}%
{\FRAC{1}{d^{\hbox to 2pt{\rm\tiny +\hss}}}{dt}}%
{\FRAC{3}{d^{\hbox to 2pt{\rm\tiny +\hss}}}{dt}}%
{\FRAC{3}{d^{\hbox to 2pt{\rm\tiny +\hss}}}{dt}}}}

\def\average#1,#2,{{1\over #2} \sum_{#1}^{#2}}

\def\eye(#1){{\bf(#1)}\quad}

\def\eq#1/{(\ref{e:#1})}

\newcommand{\beqn}[1]{\notes{#1}%
\begin{eqnarray} \elabel{#1}}

\newcommand{\eeqn}{\end{eqnarray} }

\newcommand{\beq}[1]{\notes{#1}%
\begin{equation}\elabel{#1}}

\newcommand{\eeq}{\end{equation}}

\def\bdes{\begin{description}}
\def\edes{\end{description}}

\newcounter{rmnum}

\newcounter{anum}

{\end{list}}

\def\ass(#1:#2){(#1\ref{#1:#2})}

\def\ritem#1{
\item[{\sf \ass(\current_model:#1)}]
}

\newenvironment{recall-ass}[1]{%
\begin{description}
\def\current_model{#1}}{
\end{description}
}

%% file: macros.tex
\long\def\comment#1{}

\newfont{\bb}{msbm10 scaled 1100}

\newcommand{\av}{{\bf a}}
\newcommand{\bv}{{\bf b}}

\newcommand{\ev}{{\bf e}}
\newcommand{\fv}{{\bf f}}
\newcommand{\gv}{{\bf g}}

\newcommand{\nv}{{\bf n}}

\newcommand{\qv}{{\bf q}}
\newcommand{\rv}{{\bf r}}

\newcommand{\tv}{{\bf t}}
\newcommand{\uv}{{\bf u}}
\newcommand{\wv}{{\bf w}}

\newcommand{\xv}{{\bf x}}
\newcommand{\yv}{{\bf y}}

\newcommand{\Am}{{\bf A}}

\newcommand{\Cm}{{\bf C}}

\newcommand{\Fm}{{\bf F}}

\newcommand{\Hm}{{\bf H}}
\newcommand{\Id}{{\bf I}}

\newcommand{\Um}{{\bf U}}

\newcommand{\gammav}{\hbox{\boldmath$\gamma$}}

\newcommand{\lambdav}{\hbox{\boldmath$\lambda$}}

\newcommand{\muv}{\hbox{\boldmath$\mu$}}

\newcommand{\psiv}{\hbox{\boldmath$\psi$}}

\newcommand{\rhov}{\hbox{\boldmath$\rho$}}

\newcommand{\Gammam}{\hbox{\boldmath$\Gamma$}}
\newcommand{\Lambdam}{\hbox{\boldmath$\Lambda$}}

\newcommand{\Sigmam}{\hbox{\boldmath$\Sigma$}}

\newcommand{\Psim}{\hbox{\boldmath$\Psi$}}

\renewcommand{\det}{{\hbox{det}}}

\renewcommand{\arg}{{\hbox{arg}}}

\newcommand{\herm}{{\sf H}}

\newcommand{\transp}{{\sf T}}